\documentclass{article}

\usepackage{arxiv}

\usepackage[utf8]{inputenc} 
\usepackage[T1]{fontenc}    
\usepackage{hyperref}       
\usepackage{url}            
\usepackage{booktabs}       
\usepackage{amsfonts}       
\usepackage{nicefrac}       
\usepackage{microtype}      
\usepackage{lipsum}
\usepackage{graphicx}
\graphicspath{ {./images/} }

\usepackage[greek,english]{babel}
\usepackage{geometry}

\usepackage[
singlelinecheck=false 
]{caption}

\title{EmoGator: A new open source vocal burst dataset with baseline machine learning classification methodologies}

\author{
 Fred W. Buhl \\
 University of Florida\\
  \texttt{fredbuhl@ufl.edu} \\
  }
  
\begin{document}
\maketitle
\begin{abstract}
\textit{\textbf{Vocal Bursts}} -- short, non-speech vocalizations that convey emotions, such as laughter, cries, sighs, moans, and groans -- are an often-overlooked aspect of speech emotion recognition, but an important aspect of human vocal communication. One barrier to study of these interesting vocalizations is a lack of large datasets. I am pleased to introduce the EmoGator dataset, which consists of 32,130 samples from 357 speakers, 16.9654 hours of audio; each sample classified into one of 30 distinct emotion categories by the speaker. Several different approaches to construct classifiers to identify emotion categories will be discussed, and directions for future research will be suggested. Data set is available for download from \url{https://github.com/fredbuhl/EmoGator}.
\end{abstract}



\section{Introduction}

Emotions are central to human experience---they motivate \& inform much of what we do. Recognizing emotions in others has been a longstanding area of interest. Perhaps the first scientific study of emotion recognition was the work of Duchenne \cite{duchenne_mechanism_1990} in 1862, who collected photographs of facial expressions elicited via electrically stimulating facial muscles.

The question of how many emotions there are remains open. Duchenne identified 13 primary emotions, and 60 combinations, from facial expression. A recent study by Cowen \& Keltner found that humans were able to reliably identify 28 distinct emotions from facial expression \cite{cowen_what_2019}. Another recent study by the same team \cite{cowen_self-report_2017} indicated that humans self-report as many 27 distinct emotions; these responses were collected from subjects reacting to short video clips. The emotion categories presented as gradients, which occasionally overlapped with other emotion categories; multiple emotions were elicited to varying degrees by a given stimulus.\par

Humans often express emotion vocally by varying speech \textbf{prosody}---the audio characteristics of speech. One study \cite{cowen_primacy_2019} found that 12 distinct emotions could be recognized from speech prosody---and this across two cultures---a previous study \cite{laukka_expression_2016} had found cross-cultural emotion recognition with subjects across five nations, although an in-group advantage was noted.\par

Humans also express emotion via brief, non-speech sounds called \textbf{vocal bursts}, also referred to as "affect bursts",  "emotional vocalizations", or "nonverbal vocalizations"--sounds like laughter, cries, sighs, moans, and groans---vocalizations that are not speech, and likely predate it, evolutionarily speaking. In \cite{simon-thomas_voice_2009} humans were found to be able to distinguish 14 emotional states from these vocal bursts. And a recent paper \cite{cowen_mapping_2019} by Cowen, Keltner, and others showed the ability to distinguish 24 emotional states from these brief vocalizations.\par

The ability to detect and express emotion via human vocalization appears early in human development \cite{lyakso_emotion_2015,vaillant-molina_young_2013,palama_are_2018,bloom_talking_1989,wu_one-_2017}. It is important to language and social development; people who have difficulties in discerning emotions in others, due to brain injury, or conditions like Autism Spectrum Disorder, experience difficulties communicating effectively. People with auditory affective agnosia \cite{heilman_auditory_1975} cannot discern emotional cues in speech, though they can still understand words, while people afflicted with dysprosody \cite{monrad-krohn_dysprosody_1947} speak in a monotone, without intonation or emotional affect; this can also appear in people with Parkinson's disease \cite{skodda_progression_2009}. Any impairment of these abilities has a severe effect on communication and socialization with others, underlining the importance of evoking and understanding emotional expression.\par

\subsection{The Problem at Hand}
\label{sec:1}
Interactions with computers via speech recognition is now commonplace via ``smart speakers'' and their associated virtual assistants such as Siri, Alexa, and Google Assistant. Currently, none of these systems are capable of detecting emotion from the speech audio signal; the signal is converted to text (sometimes with comic results) via speech-to-text deep learning models, but any emotional content present in the speech's prosody is ignored. For some applications, where \textit{how} a word is said may be as important (or more important) than \textit{what} word was said, this could be a severe limitation. And, given their non-speech nature, vocal bursts are completely ignored by these systems.\par

Computers capable of emotion recognition from speech have numerous applications; more life-like responses from non-player characters in video games, for example. In early childhood education, awareness of the young user's emotional state would be helpful to gauge interest, frustration, or boredom; they could also be used to assess and improve the child's emotional intelligence (or "EQ")  \cite{tsai_employing_2019}. The ability to detect emotion could detect signs of loneliness, agitation, or depression \cite{lee_diagnosis_2022}, a special concern for isolated people, such as aging-in-place seniors. Social Robots---robots designed to interact closely with humans---benefit from emotion recognition \cite{breazeal_emotion_2003}; such systems can even be used to gauge the robot's appeal to its human users \cite{novikova_sympathy_2017}. The argument has been made that we will never claim human level performance in speech recognition until we can achieve human-level speech emotion recognition, since humans are capable of both \cite{oshaughnessy_speech_2000}. (It should be noted that this area is just one aspect of the larger field of Affective Computing pioneered by Rosalind Picard \cite{picard_affective_2000}, which involve not only emotion recognition, but also emotional expression, and emotionally-aware decision making.)\par

Despite the limitations of current commercial products, Speech Emotion Recognition (SER) is an area of longstanding interest in computer science \cite{koolagudi_emotion_2012}. In 1996, Cowie et al. \cite{cowie_automatic_1996} developed a technique of automatically detecting landmarks in a speech signal and collect summary statistics, which were then used to quantify speech characteristics for four emotion categories. Various approaches have been used in speech emotion recognition over the years \cite{gadikar_survey_2020}---Mel-Frequency Cepstrum Coefficients (MFCC), Gaussian Mixture Models (GMM), Support Vector Machines (SVM), Hidden Markov Models (HMM), and neural network techniques such as LSTM \cite{hochreiter_long_1997} and, more recently, deep learning neural networks have been used.\par

The research described here examines the largely-neglected area of vocal bursts, enabled by a newly-collected dataset. A number of machine learning techniques will be explored, with varying levels of performance, along with suggested directions for future research. \par
The primary inspiration for this work was \cite{cowen_mapping_2019}; the vocal burst dataset, which the authors graciously provide to other researchers, was the largest vocal burst dataset available when released. That dataset consisted of 2,032 vocal burst samples with 30 emotion categories; as mentioned, humans were able to reliably distinguish 24 categories. The fundamental question at the basis of this current work: if humans can distinguish 24 emotion categories from vocal bursts, can machines do so as well?\par

While the Cowen et al. dataset was the largest available at the time, it was still relatively small, and the categories were not evenly represented; most machine learning approaches benefit greatly from larger numbers of samples, and balanced categories. This author determined that a larger dataset would need to be collected, and several different approaches evaluated, to find the best-performing emotion classifier.\par

\section{The dataset, and a spectrum of deep learning and other methodologies for classification}

\subsection{The Dataset}
The EmoGator dataset consists of 32,130 vocal bursts, produced by 357 speakers, providing 16.9654 hours of audio; average sample length is 1.901 seconds. Each speaker recorded three samples for each of 30 emotion categories, providing 90 samples per speaker--this provided for an equal number of samples for each category, and for each speaker, assuring equal representation in the dataset. The emotion categories were the same 30 categories used in \cite{cowen_mapping_2019}: \textit{Adoration, Amusement, Anger, Awe, Confusion, Contempt, Contentment, Desire, Disappointment, Disgust, Distress, Ecstasy, Elation, Embarrassment, Fear, Guilt, Interest, Neutral, Pain, Pride, Realization, Relief, Romantic Love, Sadness, Serenity, Shame, Surprise (Negative) Surprise (Positive), Sympathy, and Triumph}. The speakers were provided text prompts with scenarios to help elicit the emotional response; the prompts used were a modified and expanded version used by \cite{cowen_mapping_2019}, and listed in the online supplemental materials\footnote{\url{https://supp.apa.org/psycarticles/supplemental/amp0000399/amp0000399_Supplemental-Materials.docx}}.\par

Data was collected from unpaid volunteers, and also crowd-sourced workers via Mechanical Turk; a website was created where speakers could record and play back their samples using their own computer or mobile device.\par

The audio files were originally recorded at 44100 or 48000 Hz, depending on the participant's hardware, and stored as mp3 files. Each individual recording file is named with a six-digit non-sequential user id, a two-digit emotion ID (1-30), and a single-digit recording number (1,2,3). Since the files are labeled by user ID, researchers can break any train, test, or validation set by speaker, ensuring a given speaker's submission appears in only in one of the sets. (Efforts were taken to avoid a speaker providing more than one contribution, though this cannot be 100\% guaranteed). All participants provided informed consent, and all aspects of the study procedures and design were approved by the University of Florida's Institutional Review Board (IRB).\par

Quality assurance was a major part of the data collection process; there were entire submissions that were silent recordings, or only contained random background noise. Some contributors apparently misunderstood the assignment, recording themselves reading the names of the categories, or phrases related to the categories. Many speakers provided a large number of high quality samples, but also submitted problematic ones, usually due to audio issues such as background noises (for example, phone chimes or background traffic sounds); another issue was excessive breath noise picked up on the microphone. In these instances, speakers would be asked to re-record the problematic samples in order to maintain the same number of samples per speaker.\par

In addition, some speakers did not seem to be able to produce evocative speech from the prompts; their responses didn't convey distinct emotions. This last group was omitted from the dataset. As a result of all these factors, this dataset will therefore almost certainly have a bias toward the emotional expressions of North American English-speaking people, as the author, and sole evaluator, shares that personal history.\par
The dataset will be publicly available at the following URL: \url{https://github.com/fredbuhl/EmoGator}.\par

Several different steps were evaluated to preprocess the data. Normalizing the data so the range of each audio sample was within a [-1,1] range was universally used (for training, validation and testing). Denoising audio files and trimming silence from the beginning and end of audio files was evaluated as well. Augmenting data by creating pitch and time shifted variants of each sample was also explored.

While this dataset was being collected, a company named Hume AI collected their own vocal burst dataset, a subset of which was made available for the The ICML 2022 Expressive Vocalizations Workshop and Competition\cite{baird_icml_2022} as the Hume-VB dataset. This dataset consists of 59,201 vocalizations from 1702 speakers, with 10 emotion categories (\textit{Amusement, Awe, Awkwardness, Distress, Excitement, Fear, Horror, Sadness, Surprise, and Triumph}). Each sample has been rated by reviewers, with [0:100] intensity scores for every emotion category provided for each sample. This Hume-VB dataset was also used for the ACII 2022 Affective Vocal Bursts Workshop and Competition\cite{baird_acii_2022}
\par
There are several differences between the EmoGator dataset to Hume-VB dataset:
\begin{enumerate}
    \item EmoGator has 30 distinct emotion categories, with each sample belonging to a single category determined by the speaker's intent. Hume-VB has 0-100 ratings for all 10 of its categories provided by reviewers for each sample--the listener's interpretation, which may in some cases be very different than the speaker's intent.
    \item EmoGator contributors were provided text prompts describing situations that would elicit a given category of vocal burst. Hume-VB contributors were provided `seed' vocal burst audio samples to imitate--which could reduce the range of expression for a given category. 
    \item EmoGator only permitted one 90-sample submission per speaker; Hume-VB allowed for multiple submissions per speaker.
    \item EmoGator has balanced categories; each emotion category has exactly 1,071 samples. In Hume-VB, this varies; for example, ``there are fewer samples that differentially convey \textit{Triumph}'' \cite[p.~2]{baird_icml_2022}
    \item While Hume-VB has nearly twice as many samples as EmoGator, the dataset is only provided for use in the two sponsored competitions, and requires signing an End User License Agreement (EULA)\footnote{\url{https://www.competitions.hume.ai/exvo2022}}; EmoGator is freely available under an open-source license.
\end{enumerate}
At time of publication, EmoGator appears to be the largest vocal burst dataset publicly available.

\subsection{Classification Methodologies}
\label{sec:2}
A number of different techniques used in speech emotion recognition, sound classification, and elsewhere have been used for these sorts of audio classification problems.


\subsection{Spectrogram approaches}
Some approaches to audio classification involve creating a time-frequency spectrogram (or spectrogram-like) representation of the audio signals, which can be created a number of ways. Typically, the Short-Time Fourier Transform, or STFT \cite{jacobsen_sliding_2003} is used, which provides the amplitude of different frequencies over time; a variant, the Mel spectrogram, modifies the frequencies to correspond to the Mel scale \cite{stevens_scale_1937}, which closely matches human perception of differences in pitch. MFCC provide a spectrum-like ``cepstrum'' \cite{bogert_quefrency_1963}, which, while using Mel frequencies, provides the log of the amplitude in decibels over the phase shift, instead of the time domain used for spectrograms. The resulting spectrograms or cepstrograms are used as features for other machine learning approaches.\par

\subsection{1D CNN training on raw waveforms}
\label{subsec:4.2}
In \cite{dai_very_2016}, Dai et al. use a direct approach to sound classification; one-dimensional CNNs that work with the raw input waveforms, without using spectograms or some other representation as an intermediate-step feature detector. networks consisting of layers of one-dimensional convolutional neural networks (1D CNNs) \cite{kiranyaz_convolutional_2015}  were used for this. \cite{dai_very_2016} worked on the UrbanSound8k dataset \cite{salamon_dataset_2014}, which, with its 10 categories and 8,732 samples, is a bit smaller than the EmoGator dataset. Testing various architectures, they reported up to 71.68\% accuracy on an 18-layer model, which is competitive with CNNs using spectrograms of the same dataset. For the EmoGator, dataset, we developed an 18-layer network as in \cite{dai_very_2016}, and added dropout layers after each 1D convolution to help prevent overfitting.

\subsection{Random forests}
Random forest classifiers \cite{breiman_random_2001} were also explored. A random forest is constructed by generating multiple random decision trees, each constructed from a random subset of the dataset, using a random subset of each sample's features. Once constructed, each tree in the forest casts a single vote for a class, and the class with the most votes chosen the winner. This approach can be used on raw data or with spectrogram-like representations.\par

\subsection{Large pre-trained speech models} 
Several teams in the 2022 ICML Expressive Vocalizations Workshop and Competition made use of large pre-trained speech models \cite{xin_exploring_2022}, \cite{hsu_synthesizing_2022}, \cite{belanich_multitask_2022}, \cite{sharma_self-supervision_2022},\cite{purohit_comparing_2022},\cite{anuchitanukul_burst2vec_2022}. Two models were used frequently: WavLM \cite{chen_wavlm_2022} and HuBERT \cite{hsu_hubert_2021}. Both of these are self-supervised speech representation models, which are built using transformer architectures \cite{vaswani_attention_2017}; transformers have been applied successfully to a large number of domains--they are typically very large models, which have been trained on large datasets for significant amounts of time. Having access to these pre-trained models can produce better results then can be achieved by training other (usually smaller) datasets in isolation.\par

WavLM is a large scale self-supervised pre-trained speech model--The ``Large'' version of WavLM was trained on 94k hours of speech, and has 316.62M parameters. HuBERT is a similar model, the ``large'' version has 317M parameters, and was trained on 60k hours of audio on 128 Graphic Processing Units (GPUs). Both WavLM and HuBERT are built upon wav2vec 2.0 \cite{baevski_wav2vec_2020}, a ``contrastive learning'' self-supervised speech model, which itself is trained on 64 GPUs; the output of wav2vec is used as the input to HuBERT or WavLM, providing them higher-level features to build and train upon.\par

WavLM experiments were run by first running the EmoGator training, validation, and test data through a pre-trained WavLM model, storing the last hidden layer as a new representation for each sample, using a 70\% / 15\% / 15\% train-validation-test split. The hidden layers from the training data were then used as input to train a single fully connected network, using validation data to find the appropriate stopping point; once the ideal models were determined, they were run on the test data. The HuBERT model was used in a identical fashion--using the last hidden later of the HuBERT model instead of WavLM as the input to the fully-connected layer.\par

Incorporating WavLM and HuBERT in this work was greatly aided by the HuggingFace transformer libraries \cite{wolf_huggingfaces_2019}, which, while initially covering natural language processing, have now expanded into many other areas. The benefit of being able to incorporate an large pre-trained language model with a few lines of code cannot be overstated.\par

\subsection{Ensemble Methods}
Ensemble methods attempt to improve performance by combining the outputs of multiple models, with suitable training and weighting; the aggregate often outperforms the individual models. Two approaches were used for the EmoGator data: \textbf{Ensemble A} took the n-length output (where n was the number of emotion categories) produced by the WavLM-and-HuBERT-single-layer model and averaged them together, using the resulting average to pick the most likely emotion category. \textbf{Ensemble B} concatenated the last hidden layers from WavLM and HuBERT, and then trained single fully-connected layer on those inputs.

\subsection{Platform \& Hardware Requirements}
Most work on this project was performed on the University of Florida's HiperGator-AI cluster, which uses 80G A100 GPUs; one A100 should be sufficient to run all the models included, but the code may not run directly on systems with lower memory GPUs unless modifications to parameters such as batch size etc. are implemented.\par


\section{3. Results}
\subsection{1D CNN training on raw waveforms}
\begin{figure}[ht]
\includegraphics[width=15cm]{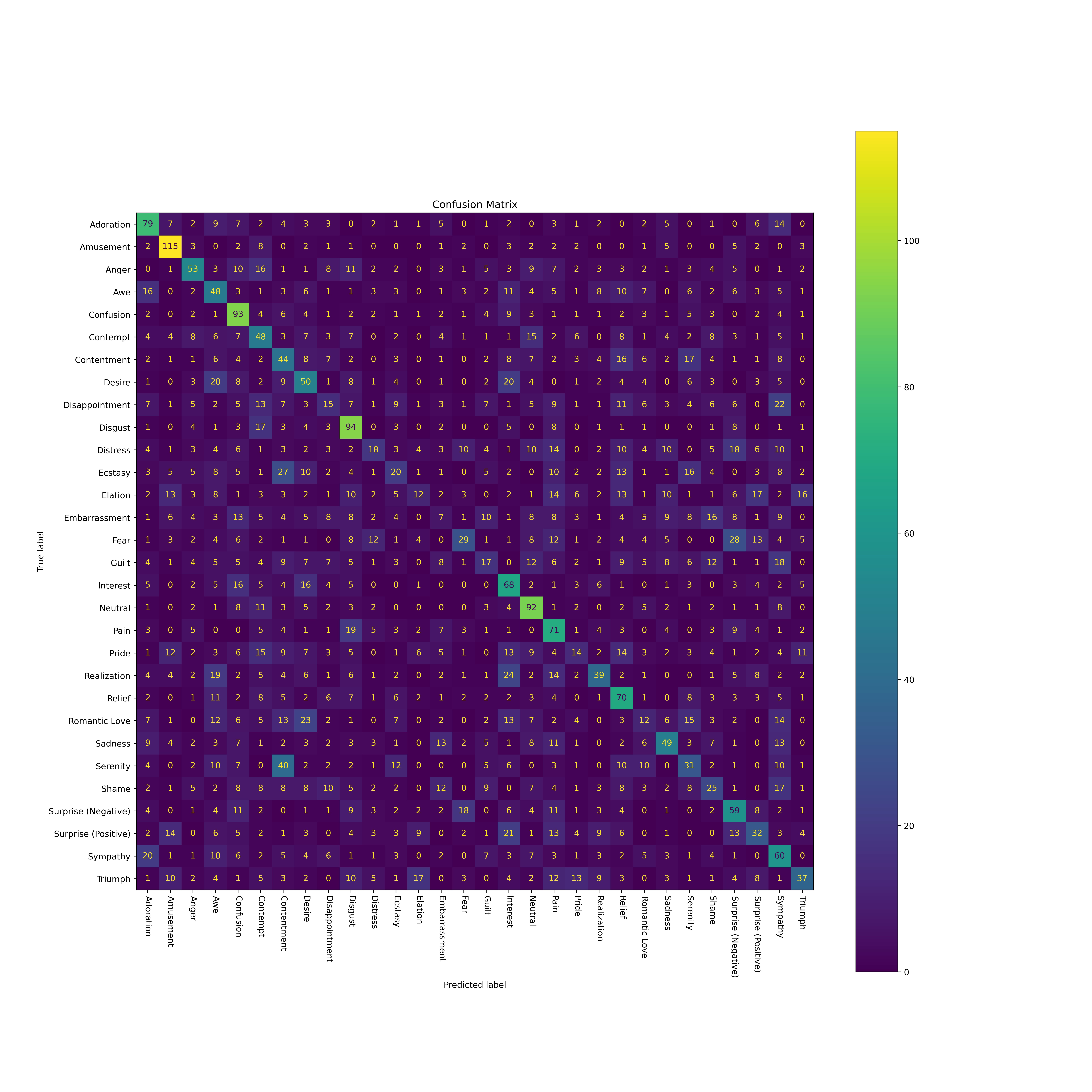}
\caption{The confusion matrix generated by the 18 layer 1D CNN with dropout layers.\label{fig1}}
\end{figure}

For one-dimensional convolutional neural networks, the best results against the full dataset were with a 70\% / 15\% / 15\% train/validation/test split, using an 18-layer 1D CNN based on \cite{dai_very_2016}, but with dropout layers after each convolution. A relatively low dropout rate of 0.07 was optimal. All experiments were run with a batchsize of 128 and an Adam optimizer with a learning rate of 0.001. Several statistics were calculated; For the full 30-category dataset, the average F1 score was 0.270. F1 scores and other accuracy metrics, with breakdowns by category, are shown in Table~\ref{tab1}; a confusion matrix is provided in Figure~\ref{fig1} based on the run with the highest F1 score. \par 
The experiments above were all run with normalized audio data, but without denoising the audio signal or trimming silence from the beginning and end; earlier experiments with a 70\%/30\% train/test split revealed that denoising or trimming the audio signal reduced performance.\par
Data augmentation was also explored; two-to-three times larger ``stretched'' version of the 70\% / 15\% / 15\% training set were produced by creating new samples by performing independent pitch and tempo shifts of the audio samples; however the stretched training sets produced lower performance than the original training set, despite making adjustments to the amount of pitch and tempo scaling.\par

In reviewing these results, it is clear that some categories are much harder (or easier) to identify; for example, the F1 score (0.056) for Embarrassment, the worst performing category, is much lower than the highest performing category, Amusement (0.627). The confusion matrix illustrates the problem well; it shows that certain types of vocal bursts are simply difficult to place in the correct category. Per the confusion matrix, Embarrassment (with only 7 samples correctly identified) was more likely to be interpreted as Shame (16) or Guilt (10); all closely related concepts that can produce similar vocalizations. This is an inherently difficult problem, which helps explain why humans could only reliably distinguish 24 emotion categories in \cite{cowen_mapping_2019}.\par

By selectively removing emotion categories that performed poorly, it would be expected that overall performance should improve. Using the F1 score as a metric, the lowest scoring categories were removed, creating 24-count, 16-count, and 10-count subsets of the dataset. Interestingly, three of the bottom-scoring six categories removed to make the 24-count subset were also not identifiable by humans in \cite{cowen_mapping_2019}; two other categories unidentifiable by humans were removed in the 16-count subset--showing some commonality between the two datasets, and also illustrating the difficulties humans and algorithms have with certain emotion categories, even across studies.\par

The same 1D CNN model architecture, hyperparameters, and validation approaches were used. Results are in Table~\ref{tab2}; we do see improvement as the more ambiguous categories are eliminated.\par

\begin{figure}[ht]
\includegraphics[width=12cm]{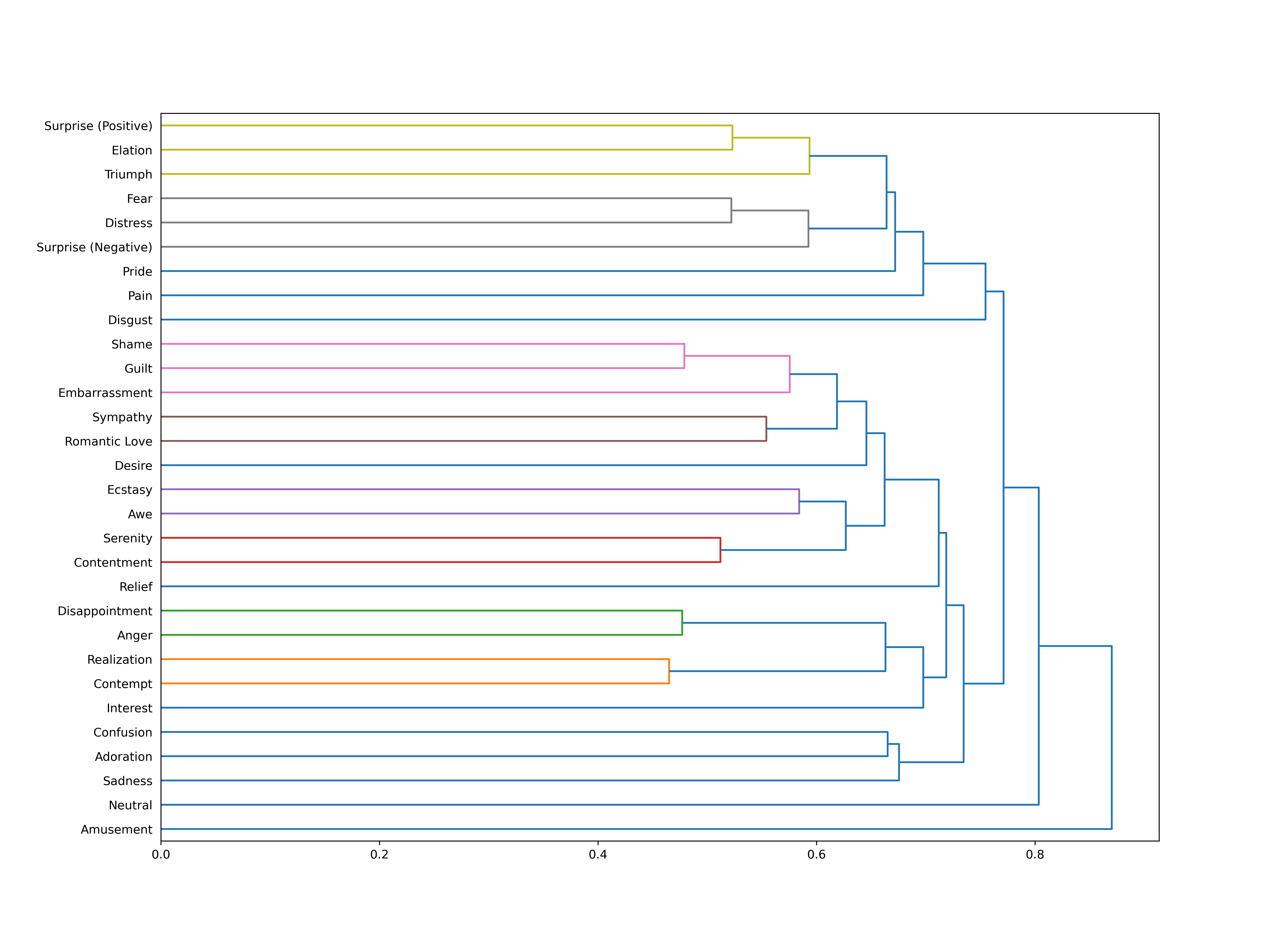}
\caption{The dendrogram generated from F1 scores (range [0,1]) between pairs of emotion categories.\label{fig2}}
\end{figure}

By creating binary 1D CNN classifiers, with one classifier for each possible pair of emotion categories, we can illustrate which pairs are the easiest to distinguish. Using the same model architecture and 70\%/15\%/15\% split, and using the F1 score as a similarity metric (on a [0,1] scale, where 1 is least similar), a similarity matrix was created based on the 435 permutations for the 30 categories, and a dendrogram displaying relationships between each category was generated from that matrix (Figure~\ref{fig2}). The dendrogram illustrates the most easily confused or distinguished categories. For example, it shows how easily the Amusement category is distinguished from all other categories, and shows Realization and Contempt as the most similar--and therefore most confused--categories, despite being very different emotions.

\begin{table}
\caption{Precision, Recall, and F1 scores from a best run of the 18 layer 1D CNN, with dropout layers.\label{tab1}}
		\begin{tabular}{rccccc}
		\toprule
& \textbf{Precision} & \textbf{Recall} & \textbf{F1 score} &  \textbf{Support}\\
\midrule
          \textbf{Adoration}    &  0.407 &    0.488    & 0.444 &      162 & \\
          \textbf{Amusement}    &  0.561 &    0.710 &    0.627  &     162 & \\
              \textbf{Anger}    &  0.405 &    0.327 &    0.362  &     162 & \\
                \textbf{Awe}    &  0.220 &    0.296 &    0.253  &     162 & \\
          \textbf{Confusion}    &  0.354 &    0.574 &    0.438  &     162 & \\
           \textbf{Contempt}    &  0.236 &    0.296 &    0.263  &     162 & \\
        \textbf{Contentment}    &  0.193 &    0.272 &    0.226  &     162 & \\
             \textbf{Desire}    &  0.253 &    0.309 &    0.278  &     162 & \\
     \textbf{Disappointment}    &  0.144 &    0.093 &    0.113  &     162 & \\
            \textbf{Disgust}    &  0.376 &    0.580 &    0.456  &     162 & \\
           \textbf{Distress}    &  0.243 &    0.111 &    0.153  &     162 & \\
            \textbf{Ecstasy}    &  0.187 &    0.123 &    0.149  &     162 & \\
            \textbf{Elation}    &  0.190 &    0.074 &    0.107  &     162 & \\
      \textbf{Embarrassment}    &  0.078 &    0.043 &    0.056  &     162 & \\
               \textbf{Fear}    &  0.341 &    0.179 &    0.235  &     162 & \\
              \textbf{Guilt}    &  0.175 &    0.105 &    0.131  &     162 & \\
           \textbf{Interest}    &  0.288 &    0.420 &    0.342  &     162 & \\
            \textbf{Neutral}    &  0.397 &    0.568 &    0.467  &     162 & \\
               \textbf{Pain}    &  0.276 &    0.438 &    0.339  &     162 & \\
              \textbf{Pride}    &  0.175 &    0.086 &    0.116  &     162 & \\
        \textbf{Realization}    &  0.351 &    0.241 &    0.286  &     162 & \\
             \textbf{Relief}    &  0.294 &    0.432 &    0.350  &     162 & \\
      \textbf{Romantic Love}    &  0.121 &    0.074 &    0.092  &     162 & \\
            \textbf{Sadness}    &  0.355 &    0.302 &    0.327  &     162 & \\
           \textbf{Serenity}    &  0.209 &    0.191 &    0.200  &     162 & \\
              \textbf{Shame}    &  0.197 &    0.154 &    0.173  &     162 & \\
\textbf{Surprise (Negative)}    &  0.296 &    0.364 &    0.327  &     162 & \\
\textbf{Surprise (Positive)}    &  0.248 &    0.198 &    0.220  &     162 & \\
           \textbf{Sympathy}    &  0.233 &    0.370 &    0.286  &     162 & \\
            \textbf{Triumph}    &  0.378 &    0.228 &    0.285  &     162 & \\
\midrule
           \textbf{Accuracy}    &       &           &    0.288 &     4860 & \\
          \textbf{Macro Average}    &  0.273    & 0.288 &    0.270 &     4860 & \\
       \textbf{Weighted Average}     & 0.273&     0.288   &  0.270 &     4860 & \\
       \bottomrule
\end{tabular}
\end{table}

\begin{table}
\caption{1D CNN runs with 24, 16, and 10 category subsets of the EmoGator dataset, compared to the 30 category full dataset.\label{tab2}}
		\begin{tabular}{lc}
		\toprule
\textbf{1D CNN Dataset size}  & \textbf{F1 score (avg.)} \\
30-Count Full Dataset & 0.267 \\
24-Count Subset & 0.344 \\
16-Count Subset & 0.459 \\
10-Count Subset & 0.597 \\
\bottomrule
\end{tabular}
\end{table}

\subsection{Random Forests}
As shown in \cite{breiman_random_2001}, an approach known as Random Forests has been used on a number of small-count, small number-of-category datasets, which suggested it might be an apt choice for the EmoGator dataset. The classifier (which is included in the scikit-learn library \cite{pedregosa_scikit-learn:_2011}) was trained against Mel-Frequency Cepstral Coefficients (MFCC) of the audio data; runs were completed for the full 30 category dataset, along with 24, 16, and 10 category subsets. Results all under-performed the 1D CNN results, however (see Table~\ref{tab3}).\par

\begin{table}
\caption{Random Forest runs with 24, 16, and 10 category subsets of the EmoGator dataset, compared to the 30 category full dataset, using MFCCs.\label{tab3}}
		\begin{tabular}{lc}
		\toprule
\textbf{Random Forest Dataset size}  & \textbf{F1 score (avg.)} \\
30-Count Full Dataset & 0.146 \\
24-Count Subset & 0.180 \\
16-Count Subset & 0.256 \\
10-Count Subset & 0.345 \\
\bottomrule
\end{tabular}
\end{table}

\subsection{Large pre-trained speech models} 
Results were calculated using the last hidden layer of WavLM and HuBERT models connected to a single fully-connected network layer. A variant of Ensemble B incorporated two fully-connected layers (labeled ``2-layer FC''), which resulted in a moderate improvement. These results are presented, along with others, in Table~\ref{tab4}.

\subsection{Ensemble Methods}
Results were calculated using averaged output from the trained fully-connected layers appended on WavLM and HuBERT model runs (Ensemble A), and concatenated last-hidden-layer outputs from both models (Ensemble B), which were then used to train a single fully-connected layer. The WavLM and HuBERT single fully-connected layers that had the highest average F1 scores on the \textit{validation dataset} were used to keep the test data from tainting the ensemble model.\par

Results for the Ensemble methods are presented in Table~\ref{tab4}, along with summary data from all the EmoGator experiments.

\begin{table}
\caption{All results from the various approaches and dataset subsets used.\label{tab4}}
\begin{tabular}{lcc}
\toprule
\textbf{Approach}  & \# \textbf{Categories} & \textbf{F1 score} \\
1D CNN & 30 & 0.267 \\
1D CNN & 24 & 0.344 \\
1D CNN & 16 & 0.459 \\
1D CNN & 10 & \textbf{0.597} \\
Random Forest & 30 & 0.146 \\
Random Forest & 24 & 0.180 \\
Random Forest & 16 & 0.256 \\
Random Forest & 10 & 0.345 \\
WavLM & 30 & 0.255 \\
WavLM & 10 & 0.563 \\
HuBERT & 10 & 0.531 \\
Ensemble A & 10 & 0.571 \\
Ensemble B & 10 & 0.591 \\
Ensemble B (2-layer FC) & 10 & 0.593 \\
\bottomrule
\end{tabular}
\end{table}

\section{Discussion}

Returning to our research question--whether, like humans, machines could reliably identify 24 emotion categories--it appears that the results achieved for the 24-emotion category runs did not approach assumed human proficiency, with a top F1 score of only 0.344 via the 1D CNN method on a 24-category subset. Results for the 24, 16, and 10-category subsets were better than the full 30-category runs, with the 10-category runs performing the best, again using the 1D CNN approach, scoring 0.597. (To put these results into perspective, a random guess for a 24-category subset would be right only 4.2\% of the time; a 10-category random guess would be right only 10\% of the time--so these results are much better than pure chance.)\par

One potential use of this dataset would be to use it to measure how accurate human performance is for vocal bursts--whether the category the speaker intended to convey is correctly identified by listeners. Other studies have used gradient rating scales for each category provided by the listener, without necessarily linking back to the ground truth of the speaker intent. Another question is whether collecting vocal bursts inspired by text-based prompts is better or worse than trying to capture them ``in the wild'' from recorded conversations, or elicited by other sorts of prompts.

Collecting more data would no doubt improve these results; this vocal burst dataset, while (currently) the largest publicly available, is still small by machine learning standards. Evaluating subsets of the dataset makes the situation even worse; when looking at say, 10-category subsets, only \(\frac{1}{3}\) of the dataset is used. 

Using more complex ensemble methods seems a promising way forward; while the ensemble results here did not exceed the 1D CNN results, it's possible that incorporating more individual models could increase accuracy beyond what we've been able to achieve. 

One topic that was not explored here is \textit{generating} vocal bursts; the author will be next exploring methods such as Generative Adversarial Networks (GANs) and Stable Diffusion models to generate vocal bursts; ideally these could be tailored for an individual speaker by providing a few audio samples(the ICML competition had this as one of their challenges).

More data will help, but it may be that audio data alone will be insufficient to properly classify vocal bursts. Datasets and models incorporating video as well as audio data--not only to look at facial expressions, but also any visual cues that might evoke a vocal burst--could improve accuracy. The words spoken by the utterer, and others around them, before or after a vocal burst may also aid in identification. (It may be, however, that there are inherent limits far short of certainty for vocal burst classification, regardless of any additional information that can be gathered--often cries of sadness and amusement sound the same, and people sometimes say they are not sure ``whether they should laugh or cry''.)
\par

Another area to explore are the demographics of the speakers; their age, gender, place of origin, and cultural background could all come into play on classifying bursts. These demographic concerns also extend to the person evaluating the quality of the sample; ideally, the demographic aspects of the reviewer should match those of the submitter for best quality.

Beyond the demographic aspects, each individual's unique character and personality certainly comes into play when they generative vocal bursts--so prior experience with the utterer could be key in improving accuracy, especially if the model's weights can be fine-tuned based on these experiences.\par

It is hoped that the EmoGator dataset will be introduce researchers to the fascinating area of vocal bursts; hopefully other researchers could incorporate this dataset into still-larger collections in the future, ``paying it forward'' by making those datasets publicly available.\par

\section*{Acknowledgement}
My thanks to Anand Rangarajan for our helpful discussions about the project.

\bibliographystyle{unsrt}  
\bibliography{references}

\begin{thebibliography}{10}

\bibitem{duchenne_mechanism_1990}
G.B. Duchenne, G.B.D. de~Boulogne, R.A. Cuthbertson, A.S.R. Manstead, and
  K.~Oatley.
\newblock {\em The {Mechanism} of {Human} {Facial} {Expression}}.
\newblock Cambridge books online. Cambridge University Press, 1990.

\bibitem{cowen_what_2019}
Alan~S. Cowen and Dacher Keltner.
\newblock What the face displays: {Mapping} 28 emotions conveyed by
  naturalistic expression.
\newblock {\em American Psychologist}, pages No Pagination Specified--No
  Pagination Specified, 2019.

\bibitem{cowen_self-report_2017}
Alan~S. Cowen and Dacher Keltner.
\newblock Self-report captures 27 distinct categories of emotion bridged by
  continuous gradients.
\newblock {\em Proceedings of the National Academy of Sciences},
  114(38):E7900--E7909, September 2017.

\bibitem{cowen_primacy_2019}
Alan~S. Cowen, Petri Laukka, Hillary~Anger Elfenbein, Runjing Liu, and Dacher
  Keltner.
\newblock The primacy of categories in the recognition of 12 emotions in speech
  prosody across two cultures.
\newblock {\em Nature Human Behaviour}, 3(4):369--382, April 2019.

\bibitem{laukka_expression_2016}
Petri Laukka, Hillary~Anger Elfenbein, Nutankumar~S. Thingujam, Thomas
  Rockstuhl, Frederick~K. Iraki, Wanda Chui, and Jean Althoff.
\newblock The expression and recognition of emotions in the voice across five
  nations: {A} lens model analysis based on acoustic features.
\newblock {\em Journal of Personality and Social Psychology}, 111(5):686--705,
  November 2016.

\bibitem{simon-thomas_voice_2009}
Emiliana~R. Simon-Thomas, Dacher~J. Keltner, Disa Sauter, Lara Sinicropi-Yao,
  and Anna Abramson.
\newblock The voice conveys specific emotions: {Evidence} from vocal burst
  displays.
\newblock {\em Emotion}, 9(6):838--846, 2009.

\bibitem{cowen_mapping_2019}
Alan~S. Cowen, Hillary~Anger Elfenbein, Petri Laukka, and Petri Keltner.
\newblock Mapping 24 emotions conveyed by brief human vocalization.
\newblock {\em American Psychologist}, 74(6):698, 2019.

\bibitem{lyakso_emotion_2015}
Elena Lyakso and Olga Frolova.
\newblock Emotion {State} {Manifestation} in {Voice} {Features}: {Chimpanzees},
  {Human} {Infants}, {Children}, {Adults}.
\newblock In Andrey Ronzhin, Rodmonga Potapova, and Nikos Fakotakis, editors,
  {\em Speech and {Computer}}, Lecture {Notes} in {Computer} {Science}, pages
  201--208, Cham, 2015. Springer International Publishing.

\bibitem{vaillant-molina_young_2013}
Mariana Vaillant-Molina, Lorraine~E. Bahrick, and Ross Flom.
\newblock Young {Infants} {Match} {Facial} and {Vocal} {Emotional}
  {Expressions} of {Other} {Infants}.
\newblock {\em Infancy : the official journal of the International Society on
  Infant Studies}, 18(Suppl 1), August 2013.

\bibitem{palama_are_2018}
Amaya Palama, Jennifer Malsert, and Edouard Gentaz.
\newblock Are 6-month-old human infants able to transfer emotional information
  (happy or angry) from voices to faces? {An} eye-tracking study.
\newblock {\em PLOS ONE}, 13(4):e0194579, April 2018.

\bibitem{bloom_talking_1989}
Lois Bloom and Richard Beckwith.
\newblock Talking with {Feeling}: {Integrating} {Affective} and {Linguistic}
  {Expression} in {Early} {Language} {Development}.
\newblock {\em Cognition and Emotion}, 3(4):313--342, October 1989.
\newblock Publisher: Routledge \_eprint:
  https://doi.org/10.1080/02699938908412711.

\bibitem{wu_one-_2017}
Yang Wu, Paul Muentener, and Laura~E. Schulz.
\newblock One- to four-year-olds connect diverse positive emotional
  vocalizations to their probable causes.
\newblock {\em Proceedings of the National Academy of Sciences},
  114(45):11896--11901, November 2017.

\bibitem{heilman_auditory_1975}
K.~M. Heilman, R.~Scholes, and R.~T. Watson.
\newblock Auditory affective agnosia. {Disturbed} comprehension of affective
  speech.
\newblock {\em Journal of Neurology, Neurosurgery \& Psychiatry}, 38(1):69--72,
  January 1975.
\newblock Publisher: BMJ Publishing Group Ltd Section: Research Article.

\bibitem{monrad-krohn_dysprosody_1947}
G.~H. Monrad-Krohn.
\newblock Dysprosody or altered "melody of language.".
\newblock {\em Brain: A Journal of Neurology}, 70:405--415, 1947.
\newblock Place: United Kingdom Publisher: Oxford University Press.

\bibitem{skodda_progression_2009}
Sabine Skodda, Heiko Rinsche, and Uwe Schlegel.
\newblock Progression of dysprosody in {Parkinson}'s disease over time—{A}
  longitudinal study.
\newblock {\em Movement Disorders}, 24(5):716--722, 2009.
\newblock \_eprint:
  https://movementdisorders.onlinelibrary.wiley.com/doi/pdf/10.1002/mds.22430.

\bibitem{tsai_employing_2019}
Tsai-Hsuan Tsai, Hsien-Tsung Chang, Shin-Da Liao, Hui-Fang Chiu, Ko-Chun Hung,
  Chun-Yi Kuo, and Chih-Wei Yang.
\newblock Employing a {Voice}-{Based} {Emotion}-{Recognition} {Function} in a
  {Social} {Chatbot} to {Foster} {Social} and {Emotional} {Learning} {Among}
  {Preschoolers}.
\newblock In Constantine Stephanidis, editor, {\em {HCI} {International} 2019
  – {Late} {Breaking} {Papers}}, Lecture {Notes} in {Computer} {Science},
  pages 341--356, Cham, 2019. Springer International Publishing.

\bibitem{lee_diagnosis_2022}
Young-Shin Lee and Won-Hyung Park.
\newblock Diagnosis of {Depressive} {Disorder} {Model} on {Facial} {Expression}
  {Based} on {Fast} {R}-{CNN}.
\newblock {\em Diagnostics}, 12(2):317, January 2022.

\bibitem{breazeal_emotion_2003}
Cynthia Breazeal.
\newblock Emotion and sociable humanoid robots.
\newblock {\em International Journal of Human-Computer Studies},
  59(1):119--155, July 2003.

\bibitem{novikova_sympathy_2017}
Jekaterina Novikova, Christian Dondrup, Ioannis Papaioannou, and Oliver Lemon.
\newblock Sympathy {Begins} with a {Smile}, {Intelligence} {Begins} with a
  {Word}: {Use} of {Multimodal} {Features} in {Spoken} {Human}-{Robot}
  {Interaction}.
\newblock {\em arXiv:1706.02757v1 [cs]}, June 2017.

\bibitem{oshaughnessy_speech_2000}
D.~O'Shaughnessy.
\newblock {\em Speech {Communications}: {Human} and {Machine}}.
\newblock Wiley, 2000.

\bibitem{picard_affective_2000}
Rosalind~W. Picard.
\newblock Affective {Computing}.
\newblock In {\em Affective {Computing}}. The MIT Press, 2000.

\bibitem{koolagudi_emotion_2012}
Shashidhar~G. Koolagudi and K.~Sreenivasa Rao.
\newblock Emotion recognition from speech: a review.
\newblock {\em International Journal of Speech Technology}, 15(2):99--117, June
  2012.

\bibitem{cowie_automatic_1996}
R.~Cowie and E.~Douglas-Cowie.
\newblock Automatic statistical analysis of the signal and prosodic signs of
  emotion in speech.
\newblock In {\em Proceeding of {Fourth} {International} {Conference} on
  {Spoken} {Language} {Processing}. {ICSLP} '96}, volume~3, pages 1989--1992
  vol.3, October 1996.

\bibitem{gadikar_survey_2020}
Akanksha Gadikar, Omkar Gokhale, Subodh Wagh, Anjali Wankhede, and P.~Joshi.
\newblock A {Survey} on {Speech} {Emotion} {Recognition} by {Using} {Neural}
  {Networks}.
\newblock {\em International Journal of Research and Analytical Reviews}, 7(3),
  September 2020.

\bibitem{hochreiter_long_1997}
Sepp Hochreiter and Jürgen Schmidhuber.
\newblock Long {Short}-{Term} {Memory}.
\newblock {\em Neural Computation}, 9(8):1735--1780, 1997.

\bibitem{baird_icml_2022}
Alice Baird, Panagiotis Tzirakis, Gauthier Gidel, Marco Jiralerspong, Eilif~B.
  Muller, Kory Mathewson, Björn Schuller, Erik Cambria, Dacher Keltner, and
  Alan Cowen.
\newblock The {ICML} 2022 {Expressive} {Vocalizations} {Workshop} and
  {Competition}: {Recognizing}, {Generating}, and {Personalizing} {Vocal}
  {Bursts}, July 2022.
\newblock arXiv:2205.01780 [cs, eess].

\bibitem{baird_acii_2022}
Alice Baird, Panagiotis Tzirakis, Jeffrey~A. Brooks, Christopher~B. Gregory,
  Björn Schuller, Anton Batliner, Dacher Keltner, and Alan Cowen.
\newblock The {ACII} 2022 {Affective} {Vocal} {Bursts} {Workshop} \&
  {Competition}: {Understanding} a critically understudied modality of
  emotional expression, July 2022.
\newblock arXiv:2207.03572 [cs, eess].

\bibitem{jacobsen_sliding_2003}
E.~Jacobsen and R.~Lyons.
\newblock The sliding {DFT}.
\newblock {\em IEEE Signal Processing Magazine}, 20(2):74--80, March 2003.
\newblock Conference Name: IEEE Signal Processing Magazine.

\bibitem{stevens_scale_1937}
S.~S. Stevens, J.~Volkmann, and E.~B. Newman.
\newblock A {Scale} for the {Measurement} of the {Psychological} {Magnitude}
  {Pitch}.
\newblock {\em The Journal of the Acoustical Society of America},
  8(3):185--190, January 1937.
\newblock Publisher: Acoustical Society of America.

\bibitem{bogert_quefrency_1963}
B.~Bogert.
\newblock The quefrency analysis of time series for echoes : cepstrum,
  pseudo-autocovariance, cross-cepstrum and saphe cracking.
\newblock In {\em Proceedings of the {Symposium} on {Time} {Series}
  {Analysis}}, pages 209--243, 1963.

\bibitem{dai_very_2016}
Wei Dai, Chia Dai, Shuhui Qu, Juncheng Li, and Samarjit Das.
\newblock Very {Deep} {Convolutional} {Neural} {Networks} for {Raw}
  {Waveforms}.
\newblock {\em arXiv:1610.00087 [cs]}, October 2016.
\newblock arXiv: 1610.00087.

\bibitem{kiranyaz_convolutional_2015}
S.~Kiranyaz, T.~Ince, R.~Hamila, and M.~Gabbouj.
\newblock Convolutional {Neural} {Networks} for patient-specific {ECG}
  classification.
\newblock In {\em 2015 37th {Annual} {International} {Conference} of the {IEEE}
  {Engineering} in {Medicine} and {Biology} {Society} ({EMBC})}, pages
  2608--2611, August 2015.
\newblock ISSN: 1558-4615.

\bibitem{salamon_dataset_2014}
Justin Salamon, Christopher Jacoby, and Juan~Pablo Bello.
\newblock A {Dataset} and {Taxonomy} for {Urban} {Sound} {Research}.
\newblock In {\em Proceedings of the 22nd {ACM} international conference on
  {Multimedia}}, {MM} '14, pages 1041--1044, Orlando, Florida, USA, November
  2014. Association for Computing Machinery.

\bibitem{breiman_random_2001}
Leo Breiman.
\newblock Random {Forests}.
\newblock {\em Machine Learning}, 45(1):5--32, October 2001.

\bibitem{xin_exploring_2022}
Detai Xin, Shinnosuke Takamichi, and Hiroshi Saruwatari.
\newblock Exploring the {Effectiveness} of {Self}-supervised {Learning} and
  {Classifier} {Chains} in {Emotion} {Recognition} of {Nonverbal}
  {Vocalizations}, June 2022.
\newblock arXiv:2206.10695 [cs, eess].

\bibitem{hsu_synthesizing_2022}
Chin-Cheng Hsu.
\newblock Synthesizing {Personalized} {Non}-speech {Vocalization} from
  {Discrete} {Speech} {Representations}, June 2022.
\newblock arXiv:2206.12662 [cs, eess].

\bibitem{belanich_multitask_2022}
Josh Belanich, Krishna Somandepalli, Brian Eoff, and Brendan Jou.
\newblock Multitask vocal burst modeling with {ResNets} and pre-trained
  paralinguistic {Conformers}, June 2022.
\newblock arXiv:2206.12494 [cs, eess].

\bibitem{sharma_self-supervision_2022}
Roshan Sharma, Tyler Vuong, Mark Lindsey, Hira Dhamyal, Rita Singh, and Bhiksha
  Raj.
\newblock Self-supervision and {Learnable} {STRFs} for {Age}, {Emotion}, and
  {Country} {Prediction}, June 2022.
\newblock arXiv:2206.12568 [cs, eess].

\bibitem{purohit_comparing_2022}
Tilak Purohit, Imen~Ben Mahmoud, Bogdan Vlasenko, and Mathew~Magimai Doss.
\newblock Comparing supervised and self-supervised embedding for {ExVo}
  {Multi}-{Task} learning track, June 2022.
\newblock arXiv:2206.11968 [cs, eess].

\bibitem{anuchitanukul_burst2vec_2022}
Atijit Anuchitanukul and Lucia Specia.
\newblock {Burst2Vec}: {An} {Adversarial} {Multi}-{Task} {Approach} for
  {Predicting} {Emotion}, {Age}, and {Origin} from {Vocal} {Bursts}, June 2022.
\newblock arXiv:2206.12469 [cs, eess].

\bibitem{chen_wavlm_2022}
Sanyuan Chen, Chengyi Wang, Zhengyang Chen, Yu~Wu, Shujie Liu, Zhuo Chen, Jinyu
  Li, Naoyuki Kanda, Takuya Yoshioka, Xiong Xiao, Jian Wu, Long Zhou, Shuo Ren,
  Yanmin Qian, Yao Qian, Jian Wu, Michael Zeng, Xiangzhan Yu, and Furu Wei.
\newblock {WavLM}: {Large}-{Scale} {Self}-{Supervised} {Pre}-{Training} for
  {Full} {Stack} {Speech} {Processing}, June 2022.
\newblock arXiv:2110.13900 [cs, eess].

\bibitem{hsu_hubert_2021}
Wei-Ning Hsu, Benjamin Bolte, Yao-Hung~Hubert Tsai, Kushal Lakhotia, Ruslan
  Salakhutdinov, and Abdelrahman Mohamed.
\newblock {HuBERT}: {Self}-{Supervised} {Speech} {Representation} {Learning} by
  {Masked} {Prediction} of {Hidden} {Units}, June 2021.
\newblock arXiv:2106.07447 [cs, eess].

\bibitem{vaswani_attention_2017}
Ashish Vaswani, Noam Shazeer, Niki Parmar, Jakob Uszkoreit, Llion Jones,
  Aidan~N Gomez, Lukasz Kaiser, and Illia Polosukhin.
\newblock Attention is {All} you {Need}.
\newblock {\em 31st NIPS Conference Proceedings}, 2017.

\bibitem{baevski_wav2vec_2020}
Alexei Baevski, Henry Zhou, Abdelrahman Mohamed, and Michael Auli.
\newblock wav2vec 2.0: {A} {Framework} for {Self}-{Supervised} {Learning} of
  {Speech} {Representations}.
\newblock {\em arXiv:2006.11477 [cs, eess]}, June 2020.
\newblock arXiv: 2006.11477.

\bibitem{wolf_huggingfaces_2019}
Thomas Wolf, Lysandre Debut, Victor Sanh, Julien Chaumond, Clement Delangue,
  Anthony Moi, Pierric Cistac, Tim Rault, Rémi Louf, Morgan Funtowicz, and
  Jamie Brew.
\newblock {HuggingFace}'s {Transformers}: {State}-of-the-art {Natural}
  {Language} {Processing}.
\newblock {\em arXiv:1910.03771 [cs]}, October 2019.
\newblock arXiv: 1910.03771.

\bibitem{pedregosa_scikit-learn:_2011}
Fabian Pedregosa, Gaël Varoquaux, Alexandre Gramfort, Vincent Michel, Bertrand
  Thirion, Olivier Grisel, Mathieu Blondel, Peter Prettenhofer, Ron Weiss,
  Vincent Dubourg, Jake Vanderplas, Alexandre Passos, David Cournapeau,
  Matthieu Brucher, Matthieu Perrot, and Édouard Duchesnay.
\newblock Scikit-learn: {Machine} {Learning} in {Python}.
\newblock {\em J. Mach. Learn. Res.}, 12:2825--2830, November 2011.

\end{thebibliography}

\end{document}